\documentclass{appolb}
\usepackage{epsfig}
\usepackage{floatflt}
\usepackage{epsfig}
\begin{document}
\title{\bf  Cronin effect and energy conservation constraints  in pA collisions at LHC and RHIC.
\thanks{Presented at XXXIV International Symposium on Multiparticle Dynamics. }}
\author{E. Cattaruzza and D. Treleani
\address{Dipartimento di Fisica Teorica dell'Universit\`a di Trieste and INFN, Sezione di Trieste,\\Strada Costiera 11, Miramare-Grignano, I-34014 Trieste,Italy.}}
\maketitle
\begin{abstract}
We evaluate the Cronin effect in $pA$ collisions at the CERN LHC and at RHIC in the framework  of Glauber-Eikonal model of initial state multiparton interactions. Taking
carefully into account all kinematical constraints of each multi-parton interaction
process we obtain a softening of the spectrum of produced partons, improving in this way the agreement of the model with the recent measurements of $\pi^0$ production in $d+Au$ collisions at $\sqrt s=200$ AGeV.
\end{abstract}
\PACS{24.85+p, 11.80.La, 25.75.-q}
\section{Introduction}
Of particular interest in the context of hadron-nucleus collisions is the study of the Cronin effect, which consists of the nuclear modification of the transverse momentum spectrum of hadrons with respect to what is to be expected from a naive superposition of nucleon-nucleon collisions. Because of large scale of exchanged momenta the problem can be approached with pQCD methods, while
the target complex structure effects may be controlled by changing energy
and atomic mass number.
This effect can be explained in terms of multiple scattering of the projectile partons,
 being this latter induced by the high density of the nuclear target (see Ref.~\cite{Accardi: theor models} 
 for a review of theoretical models). 
With the adoption of the Glauber prescription of factorization of the overall many-parton $S$ matrix,
which can be expressed  as a convolution 
 of elementary partonic $S$ matrices, and with the decoupling of longitudinal and transverse degrees of freedom, 
 the longitudinal component of the incoming projectile parton is conserved and all multiparton interactions
 of the projectile can consequently be summed in the following analytical formula for the inclusive transverse spectrum: 
\begin{equation}
\frac{d\sigma}{d^2bdxd^2p_t}=\frac{1}{(2\pi)^2}\int d^2r\,e^{ip_tr}\,G(x)\,S_{hard}^A(\bar r,\bar b, p_0), \label{multiscattering-series}
\end{equation}
where 
\begin{eqnarray*}
S_{hard}^{A}(\bar{r},\,\bar{b},\,
p_{0})=\left[e^{T_{A}(b)\tilde\sigma_{hard}^{qN}(r,p_0)}-e^{T_{A}(b)\sigma_{hard}^{qN}(p_0)}\right]
\end{eqnarray*}
and $T_A(b)$ is the usual nuclear thickness, as a function of the
hadron-nucleus impact parameter $b$, $G(x)$ the parton number density of the projectile, as a function of the fractional momentum $x$, $p_{t}$ the transverse momentum of the final observed parton. 
The previous quantity can be expressed in terms of the dipole-nucleus hard cross section
$\tilde{\sigma}_{hard}^{qN},$ which originates from the square of the scattering amplitude and depends from the transverse size $r$ of dipole:
\begin{eqnarray*}
\tilde{\sigma}_{hard}^{qN}=\int d^2p_t \left[1-e^{-i\bar p_t\cdot \bar r}\right]\,
\frac{d\sigma^p_{hard}}{d^2p_t},
\end{eqnarray*}
where  $d\sigma^p_{hard}/d^2p_t$ is the pQCD parton-nucleon cross section, this latter depending from the parton number density of the nucleus and the elementary parton-parton cross section, which includes also the kinematical constraints.  
The infrared divergences deriving from pQCD are regularized with a cut-off $p_0$; nevertheless, as unitarity is explicitly implemented, the degree of infrared singularity of the cross section is reduced from an inverse power to a power of a logarithm of the cut-off. It is important to notice that in the low $p_t$ limit unitarity produces a suppression of the integrated parton yield and a random walk of parton to  higher $p_t,$ recovering in this way the  local isotropy in transverse space of the black disk limit, which is maximally broken in the lowest order impulse approximation. 
\section{Transverse spectrum expansion}
As Glauber-Eikonal model does not account for energy conservation, the spectrum is therefore shifted towards larger transverse momenta, this effect being emphasized as the number of rescattering grows. After implementing kinematical constraints exactly, for a given final state, an increased energy is needed for the initial projectile partons, and, as the structure functions are singular in the small $x$ limit, the initial parton flux is sizably reduced: the multi-scattering series of Eq.(\ref{multiscattering-series}) cannot be resummed anymore and the expansion in the number of rescatterings is needed:
 \begin{equation}
\frac{d\sigma}{d^2bdxd^2p_t} = \frac{d\sigma^{(1)}}{d^2bdxd^2p_t}+\frac{d\sigma^{(2)}}{d^2bdxd^2p_t}+
\frac{d\sigma^{(3)}}{d^2bdxd^2p_t}+\ldots
\end{equation}
where the first term of expansion is the single scattering term (the projectile parton interacts with a single parton of the target and viceversa), while the other contributions represent  the rescattering terms (the projectile interacts with $i=2,3,\ldots$ target partons). Most of the spectrum is nevertheless well reproduced by the first three terms of the expansion Ref.~\cite{Accardi:2001ih}. Recalling that in the Glauber-Eikonal model the three body process can be expressed as a convolution of two on shell two body interactions, it is possible to reconstruct the whole kinematics keeping as independent variables the outgoing and incoming fractional momenta, the exchanged transverse momenta and implementing energy-conservation and mass shell conditions; the following j-scattering contribution to the transverse spectrum of i-parton species is obtained: 
 \begin{eqnarray*}
\frac{d\sigma^{(j)}_i}{d^2bdyd^2p_t}&\sim & T_A(b)^j\int \prod_{k=1}^j\,d^2q_k\,dx'_k\,\Delta^{(j)}(\bar{q}_1,\ldots,\bar q_j)\,\hat{\sigma}^{j}(y,x'_1,\ldots,x'_j;\bar q_1,\ldots \bar q_j)\\
&\times& x\,f_{i/p}(x,Q_{fct})\,f_{A}(x'_1,Q_{fct})\cdots f_{A}(x'_j,Q_{fct}),
\end{eqnarray*}
where $\Delta^{(j)}(\bar{q}_1,\ldots,\bar q_j)$ are the subtractive terms deriving from unitarity implementation and $\hat{\sigma}^{j}(y,x'_1,\ldots,x'_j;\bar q_1,\ldots \bar q_j)$ are the j-scattering cross sections with exact kinematics (see Ref.\cite{Cattaruzza:2004qr}
for a detailed explanation); higher order effects in the elementary interactions are accounted by multiplying the lowest order expressions in $\alpha_s$ by the factor $k_{fct}$, while the infrared divergences are regularized by a cut-off $p_0$.
\section{Numerical Results}
\begin{figure}[!h]
\begin{center}
\includegraphics[width=9cm]{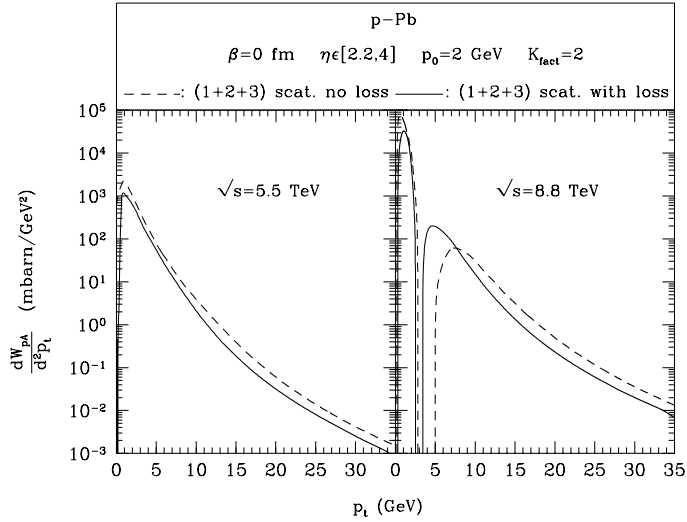}
\caption{{\small Transverse momentum spectrum of partons produced in $p+Pb$ collisions at $\sqrt{s}=5.5\, ,8.8\, ATeV\,$and $\eta\in[2.4,\,4]$, using  $p_{0}=2\, GeV,\,$ and $k_{factor}=2\,$, factorization and renormalization scales $Q_{fct},\, Q_{rn}\,$equal to the regularized transverse mass $m_{t}=\sqrt{p_{0}^{2}+p_{t}^{2}}$.}}
\label{spectra.s5500.s8800}
\end{center}
\vskip-0.9cm
\end{figure}

\begin{figure}[!t]
\includegraphics[
  width=6.30cm]{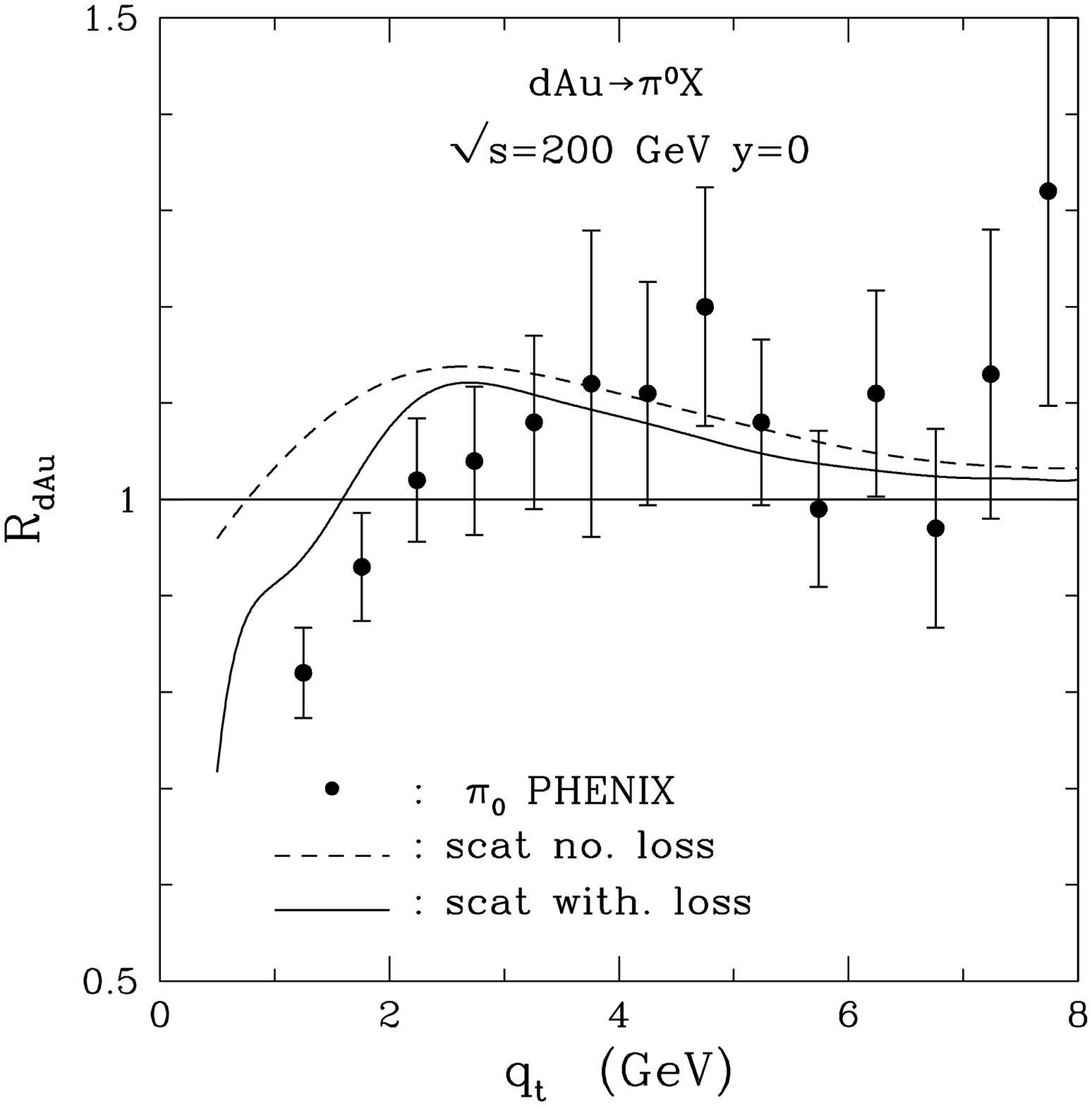}
 \includegraphics[
  width=6.2cm]{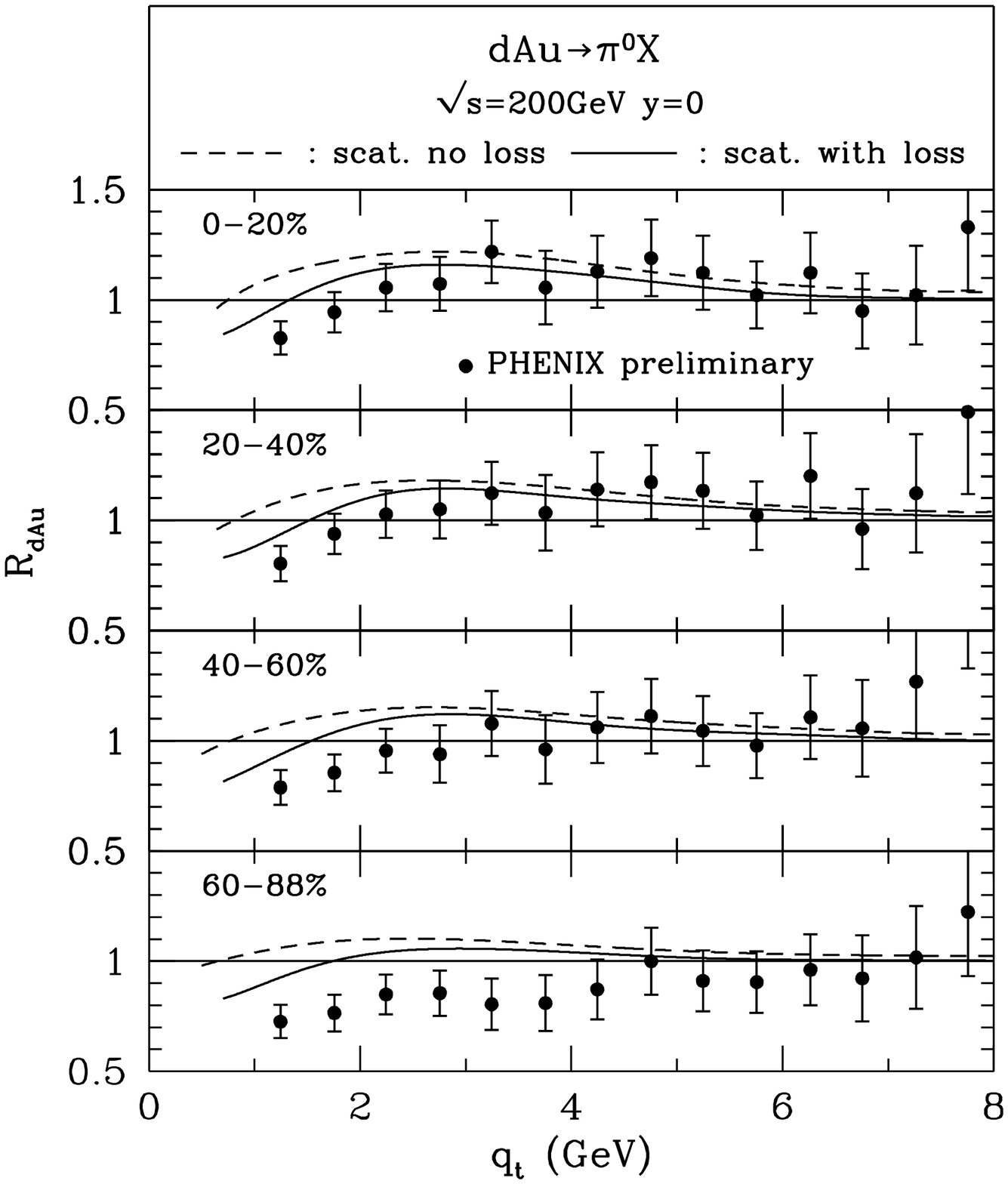}
\caption{\emph{Left Panel}: Cronin ratio in $d+Au\,\rightarrow\pi^{0}\, X$ collisions at $\sqrt{s}=200\,
GeV\,$. The solid line refers to the case of exact kinematics and the dashed line to the
case of approximate kinematics at $y=0$. The data are form Ref.\cite{Adler:2003ii}.
\emph{Right Panel}: Centrality dependence of Cronin ratio with and without energy loss implementation at $y=0$. Comparison with
experimental data presented by PHENIX.}
\label{Phenix-Centrality}
\vskip-0.58cm
\end{figure}

\begin{figure}[!t]
\begin{center}
\includegraphics[width=6.5cm]{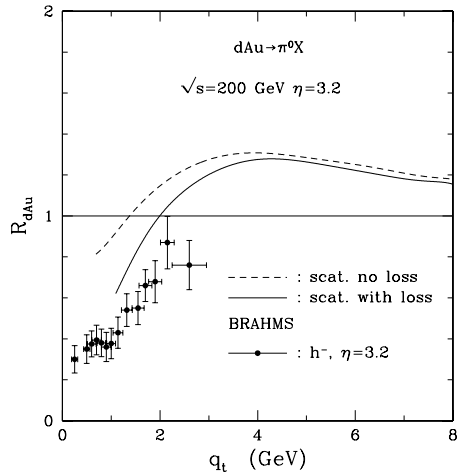}
\caption{Cronin ratio in $d+Au\,\rightarrow\pi^{0}\, X$ collisions at $\sqrt{s}=200\,
GeV\,$. The solid line refers to the case of exact kinematics and the dashed line to the
case of approximate kinematics at $\eta=3.2$. The data are form Ref.~\cite{Brahms}}
\label{Brahms}
\end{center}
\vskip -0.9cm
\end{figure}

To study the effect at the LHC we consider the case of production of minijets in a forward calorimeter ($\eta\in[2.4,4]$) at two different center of mass energies in the hadron-nucleon c.m. system $\sqrt{s}=5.5\, ,8.8\, ATeV\,$. Our results are plotted in Fig.\ref{spectra.s5500.s8800}, where we compare the spectrum obtained by the exact implementation of energy conservation in the multiple interactions (solid line), with the approximate kinematics results given by the first three terms of expansion of Eq.(\ref{multiscattering-series}).  As an effect of the exact implementation of kinematics 
the spectrum of outgoing particles is shifted toward lower transverse momenta; the entity of such a suppression is of the order of 40-50$\%$ at $p_t\sim15\,GeV,$ it is still about 30-38$\%$ for higher transverse momenta $p_t\sim30\,GeV.$ The triple scattering approximation, used to evaluate the
spectrum, breaks down at $p_t\le9\, GeV$ at $\sqrt{s}=8.8\,TeV$: by increasing the center
of mass energy the density of target partons grows rapidly and the contribution of higher
order rescatterings cannot be neglected any more at  $p_t\le9\, GeV$ (right panel of
Fig.1).
 As the energy is lowered to $\sqrt{s}=5.5\,TeV$, the
region of numerical instability is shifted to the region $p_{t}\le1\, GeV\,$(left panel
of Fig.1).
For a comparison with recent measurements of the Cronin effect in $dAu\to \pi_0X$  at RHIC, we consider the following expression for the inclusive $\pi_0$ spectrum
\begin{equation}
\frac{d\sigma^{frag}_{h}}{d^2q_t\,dy_h\,d^2b}\sim \sum_{i}\frac{d\sigma
_{i}}{d^2p_t\,dy\,d^2b}\otimes D_{i\to h}(Q_F^2),
\end{equation} 
where $D_{i\to h}(Q_F)$ are the fragmentation functions at the fragmentation scale $Q_F$. At lower energy, $\sqrt{s}=200\,$ AGeV, we follow Ref.~\cite{Accardi:2003jh} in the evaluation
of the cross section of $d+Au\rightarrow\pi^{0}X$, using
$Q_{fct}=Q_{rn}=Q_F=\frac{m_{t}}{2}\,$, and the values $p_{0}=1.0\,$ GeV and $k_{fact}=1.04$. With these choices the effects of higher order corrections in $\alpha_s$ are minimized and the inclusive cross section of $\pi_0$ production in pp collisions at the same c.m. energy is reproduced without smearing with the intrinsic $k_t$: the resulting Cronin ratio in  $d+Au\rightarrow\pi^{0}X$ is hence a parameter-free prediction of the model.
By using the leading order K-K-P fragmentation functions at $y=0$ and $b=b_{dAu}=5.7\,fm$, which is the estimated average impact parameter of the experiment, we evaluate 
\[
R_{dAu\rightarrow\pi^{0}X}=\frac{d\sigma_{dAu\rightarrow\pi^{0}X}^{frag}}{d^{2}q_{t}\,
dy\, d^{2}b}\left/\frac{d\sigma_{dAu\rightarrow\pi^{0}X}^{frag\,(1)}}{d^{2}q_{t}\, dy\,
d^{2}b}.\right.\]  
In  the left panel of Fig.\ref{Phenix-Centrality} we compare our result (continuous line) with the experimental data of Phenix collaboration Ref.~\cite{Adler:2003ii} and with the results using approximate kinematics (dashed lines). Because of the small rapidity values of the observed $\pi_0$, in this case the corrections induced by exact kinematics are important in the region of smaller transverse momenta.  
The dependence of the effect on the impact parameter is shown in the right panel of Fig.\ref{Phenix-Centrality}, where our calculation (continuous line) is compared with preliminary data presented by PHENIX at the DNP fall meeting ~\cite{DNP}, and with the standard Glauber-Eikonal calculation (dashed line).  As a consequence of exact implementation of kinematical constraints a systematical reduction of the Cronin curve is observed, improving in this way the agreement with experimental data. \\
As it can be seen from the behavior of Cronin ratio in the forward rapidity region at  $\eta=3.2$ Fig.\ref{Brahms}, the quenching  of the spectrum, due to the energy lost by the projectile in the multiple collision of the projectile, is now sizably  increased  (notice that the vertical scale is now different respect to the previous of Fig.~\ref{Phenix-Centrality});  actually this has to be attributed to the fact that as larger become the rapidity values as larger becomes the average number of rescatterings. The expectation is nevertheless that the Cronin curve should exceed one for $p_t\ge2\,GeV$; as we can see, from a comparison of our results with experimental data of BRAHMS Collaboration~\cite{Brahms}, for low $p_t$ and forward values of pseudorapidity, the behavior of the Cronin curve  can be described with a not so bad approximation also in the context of Eikonal dynamics.   
\section{Conclusions}
In high energy proton-nucleus collision the Cronin effect is successfully described by Glauber-Eikonal model, which allows the implementation of unitarity constraints. Working out the leading terms in the expansion in multiple parton collisions, we have taken into account all kinematical constraints exactly obtaining an improved agreement with the Cronin ratio measured by PHENIX in $dAu\to\pi_0X$ at $\sqrt{s}=200\,GeV$. Results at larger pseudo-rapidities indicate a reduction of the Cronin ratio, which can be partially explained by simple Eikonal dynamics.  

\end{document}